# Photoconductive detection of arbitrary polarised terahertz pulses

E. Castro-Camus, J. Lloyd-Hughes, and M.B. Johnston
*University of Oxford, Department of Physics, Clarendon Laboratory,
Parks Road, Oxford OX1 3PU, United Kingdom*
M.D. Fraser, H.H. Tan, and C. Jagadish
*Department of Electronic Materials Engineering, Research School of Physical Sciences and Engineering, Institute of Advanced Studies, Australian National University, Canberra ACT 0200, Australia*

The technique of THz time domain spectroscopy (TDS) provides a very sensitive probe across the THz band. To date THz-TDS techniques have relied on linearly polarised emitters and detectors. However, for spectroscopy of birefringent and optically active materials it is also important to measure the polarisation state of radiation. Additionally the use of circularly polarised THz radiation could enable the study of chiral molecules, such as proteins and DNA, which have vibrational and librational resonances in this part of the spectrum.

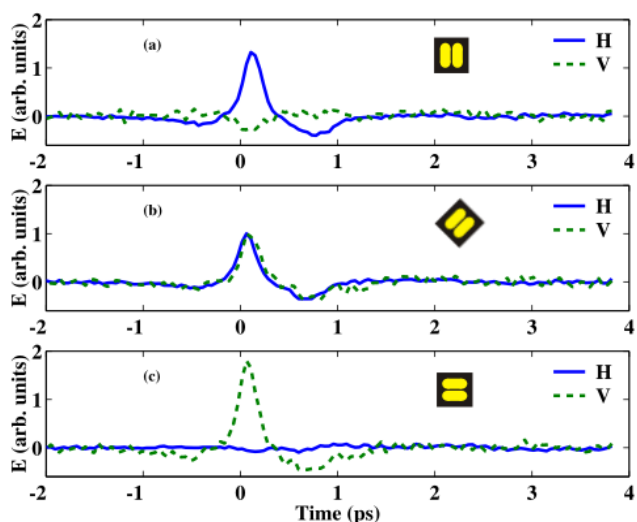

*Fig.1: The measured electric field components (H=horizontal and V= vertical) of linearly polarized THz pulses for different angles of the emitter 0, 45 and 90°.*

To perform polarization sensitive THz-TDS, it is necessary to be able to measure two (preferably orthogonal) electric field components of a terahertz transient. We have developed a polarization sensitive detector by fabricating a three-contact photoconductive receiver. In order to construct such a detector we firstly implanted an InP substrate with $Fe^+$ ions ($2.5 \times 10^{12}$ and $1.0 \times 10^{13} cm^{-2}$ at 0.8 and 2.0MeV resespectively), to give a carrier lifetime of ~130fs. Subsequently three Cr/Au contacts were deposited. The three-contact photoconductive receiver was tested using a standard TDS setup[1], linearly polarized THz transients were generated by exciting a 400 μm gap SI-GaAs photoconductive switch with 10fs duration laser pulses. Two separate lock-in amplifiers were used to record simultaneously the currents through each of the two pairs of contacts. The photoconductive emitter was mounted on a graduated rotation stage that allowed the gap, and hence the polarization of the emitted THz transient, to be rotated. The plots in Fig. 1 show both components of the THz electric field as function of time with the emitter at 0, 45 and 90° (as shown in the inset) we can se that the components at 45° are approximately equal while at 0° (90°) the vertical (horizontal) component is much smaller, this demonstrate that the three-contact photoconductive receiver acts as a polarization sensitive detector. The polarization selectivity of the detector was assessed by measuring the cross-polarised extinction ratio. This ratio was found to be 108:1 (128:1) for the horizontally (vertically) oriented emitter. This integrated three-contact detector is expected to be very useful for developing time-domain circular dichroism spectroscopy and should have a wide range of applications in basic research and industry.

[1] M.B. Johnston *et al*. Chem. Phys. Lett., **377** (2003), 256